\documentclass[a4paper,12pt]{article}
\usepackage{amsthm,amssymb,amsmath,url,amsfonts,mathrsfs,bm}
\newcommand{\FFF}{\mathfrak{F}}
\newcommand{\ops}{\overline{\psi}}

\newcommand{\bea}{\begin{eqnarray*}}
\newcommand{\eea}{\end{eqnarray*}}

\newcommand{\Pe}{Poincar\'e }

\newcommand{\Rtn}{{\mathbb{R}^{3N}}}

\newcommand{\es}{\emptyset}
\newcommand{\app}{\approx}

\newcommand{\ga}{\gamma}

\newcommand{\tit}{\textit}

\newcommand{\bg}{\begin}
\newcommand{\en}{\end}
\newcommand{\Lra}{\Leftrightarrow}
\newcommand{\tx}{\text}
\newcommand{\na}{\nabla}
\newcommand{\De}{\Delta}
\newcommand{\Si}{\Sigma}
\newcommand{\se}{\subseteq}
\newcommand{\C}{\mathbb{C}}
\newcommand{\R}{\mathbb{R}}

\newcommand{\FF}{{\cal F}}
\newcommand{\si}{\sigma}

\newcommand{\X}{\mathbb{X}}

\newcommand{\B}{\mathbb{B}}

\newcommand{\K}{\mathbb{K}}

\renewcommand{\j}{\mathbf{j}}

\usepackage{graphicx}

\begin{document}
\begin{large}
\title{\bf {Relativistic Bohmian mechanics without a preferred foliation}}
\end{large}
\author{Bruno Galvan \footnote{e-mail: b.galvan@virgilio.it}\\ \small via Melta 16, 38121 Trento, Italy.}
\maketitle
\begin{abstract}
In non-relativistic Bohmian mechanics the universe is represented by a probability space whose sample space is composed of the Bohmian trajectories. In relativistic Bohmian mechanics an entire class of empirically equivalent probability spaces can be defined, one for every foliation of spacetime. In the literature the hypothesis has been advanced that a single preferred foliation is allowed, and that this foliation derives from the universal wave function by means of a covariant law. In the present paper the opposite hypothesis is advanced, i.e., no law exists for the foliations and therefore all the foliations are allowed. The resulting model of the universe is basically the ``union'' of all the probability spaces associated with the foliations. This hypothesis is mainly motivated by the fact that any law defining a preferred foliation is empirically irrelevant. It is also argued that the absence of a preferred foliation may reduce the well known conflict between Bohmian mechanics and relativity.

\end{abstract}

\section{Introduction}

In non-relativistic Bohmian mechanics \cite{bohm,hol,dbook,dbook2,qe} the universe is represented by a probability space whose sample space is composed of the Bohmian trajectories. The actual trajectory of the universe is assumed to be a trajectory chosen at random from this space. In the relativistic domain the definition of the probability space requires selecting a foliation of spacetime; the selected foliation is however not empirically detectable, and the probability space generated by any foliation can be an empirically adequate model of the universe \cite{bohm,qe,lorentz,bemade}. The need of a foliation creates some conflict between Bohmian mechanics and relativity, even if there is not an obvious criterion which states clearly if a theory is compatible or not with relativity \cite{lorentz,hyper,bemade,maud}. In order to reduce this conflict, in \cite{bemade} the hypothesis is advanced that a single preferred foliation is allowed, and that this foliation derives from the universal wave function by means of a covariant law.

In the present paper the opposite hypothesis is advanced, namely that there is no law for the foliations, neither dynamical nor statistical. This means that all the foliations are allowed, and no probability measure is defined on the set of the foliations. This hypothesis will be synthetically referred to as the \tit{no-law hypothesis}.

The no-law hypothesis leads naturally to a model of the universe which is basically the ``union'' of the probability spaces associated with all the foliations; such a model will be referred to as the \tit{no-law model}. A remarkable feature of this model is that the random choice of a trajectory from the sample space is governed by a set function which is not a probability measure.

The main motivation for the no-law hypothesis is the fact that, due to the complete unobservability of the foliation, any law for the foliations is empirically irrelevant. However it is also argued that the absence of a preferred foliation in the no-law model may reduce the conflict between Bohmian mechanics and relativity. Both these subjects are discussed in Section \ref{disc}.

The plan of the paper is the following: in Section \ref{non} non-relativistic Bohmian mechanics and its interpretation are presented; in Section \ref{rel} relativistic Bohmian mechanics is presented; in Section \ref{nl} the no-law model is developed; in Section \ref{disc} some conceptual issues are discussed.

For simplicity, in this paper the mathematical details relative to $\si$-algebras and the measurability of sets are omitted.

\section{Non-relativistic Bohmian mechanics} \label{non}

Let us assume that the universe is a non-relativistic quantum system of $N$ particles. The configuration space of the system is $\X:=\Rtn$. Let us also introduce the \tit{kinematic space} of the system, i.e., the set $\K$ of all the kinematically admissible trajectories of the system; in this case we can define $\K:=C^0(\R; \X)$. At every time $t \in \R$ the quantum state of the system is represented by a normalized vector $\psi_t \in L^2(\X)$. The time evolution of $\psi_t$ is determined by a Hamiltonian of the form
\bg{equation}
H = - \sum_{i=1}^N \frac{\hbar^2}{2 m_i} \De_i + V(x).
\end{equation}
A probability density and a probability current can be naturally derived from the wave function:
\begin{eqnarray}
& & \rho_t(x):=|\psi_t(x)|^2, \label{nrd} \\ 
& & \j_{i t}(x):=\frac{\hbar}{m_i} \tx{Im} \,\psi_t^*(x) {\bm \na}_i \psi_t(x), \; \; i=1, \ldots, N. \label{nrf}
\end{eqnarray}
These satisfy the continuity equation
\bg{equation}
\dot \rho_t + \na \cdot j_t=0,
\end{equation}
where $j_t:=(\j_{1t}, \ldots, \j_{Nt})$. A \tit{Bohmian trajectory} is a trajectory $k \in \K$ satisfying the \tit{guiding} equation
\bg{equation} \label{diff}
\dot k(t)=\frac{j_t(k(t))}{\rho_t(k(t))}.
\end{equation}
Let $\B$ denote the set of the Bohmian trajectories. For $K \se \K$ we define $K_t:=\{k(t) \in \X: k \in K\}$. One can prove that
\bg{equation} \label{equi}
P_t(B_t)=P_{t'}(B_{t'}) \tx{ for all } B \se \B \tx{ and } t, t' \in \R,
\end{equation}
where $P_t$ is the probability measure on $\X$ induced by $\rho_t$. This property is referred to as \tit {equivariance}. It follows that the set $\B$ is naturally endowed with the probability measure
\bg{equation}
P(B):=P_t(B_t), \tx{ where } B \se \B.
\end{equation}
The probability space
\bg{equation} \label{stand}
(\B, P)
\end{equation}
will be referred to as the \tit{Bohmian space}.

\vspace{3mm}
The Bohmian space is assumed to be a model of the universe in the sense that the actual evolution of the universe is assumed to be represented by a trajectory chosen at random from $\B$.

The random choice is governed by the probability measure $P$ as follows: (1) the probability $P$ defines the typical subsets of $\B$, i.e., the sets $B \se \B$ such that $P(B) \app 1$; (2) the typical sets govern the random choice according to Cournot's principle, which states what follows: we can be practically certain that an element chosen at random from $\B$ belongs to a typical subset of $\B$ singled out in advance or independently with respect to the random choice \cite{cp,gal}.

The Bohmian space is an empirically adequate model because the (vaguely defined) set of the \tit{physical} trajectories, i.e., the trajectories which give the correct results in the experiments, is typical. This result is proved in \cite{qe}.

It has already been noted in the literature that the additive structure of a probability measure has no relevance for the definition of the typical sets and therefore these sets could be defined by non-additive set functions as well \cite{typ0,typ,gal}. This will be the case with the no-law model defined in Section \ref{nl}. Also in that case the connection between typical sets and random choice is given by Cournot's principle.

\section{Relativistic Bohmian mechanics} \label{rel}

Let us now assume that the universe is a system of $N >1$ non-interacting fermions. The formulation of relativistic Bohmian mechanics for this system is based on the hypersurface Bohm-Dirac model \cite{hyper}.

Let $M:=\R \times \R^3$ denote the Minkowski spacetime, $\Si$ a smooth space-like hypersurface of $M$ (hereafter simply hypersurface), and $\FF$ a foliation of hypersurfaces \cite{hyper}. The configuration space of the system is $\X_\Si:=\Si^N$, which in the relativistic case depends on the hypersurface. The kinematic space $\K$ is the set of the $N$-tuples of world lines\footnote{In the relativistic formulation the symbols $\K$ and $\B$ are redefined.}$^{,}$\footnote{Here a world line is considered to be a subset of $M$, and its intersection with any hypersurface is exactly one point.}. For consistency with the non-relativistic case, let us introduce the following notation: for $k=(k_1, \ldots, k_N) \in \K$ define $k(\Si):=(k_1 \cap \Si, \ldots, k_N \cap \Si) \in \X_\Si$, and for $K \se \K$ define $K_\Si:=\{k(\Si) \in \X_\Si: k \in K\}$. Note that in the relativistic case the variable $t$ is replaced by the ``variable'' $\Si$.

The wave function of the system is a multi-time function $\psi:M^N \to (\C^4)^{\otimes N}$ satisfying the system of Dirac equations ($c=\hbar=1$)
\bg{equation}
(i \ga_i \cdot \partial_i- e_i \ga_i \cdot A(x_i) - m_i)\psi = 0, \; \; i=1, \ldots, N,
\end{equation}
where $\ga_i = I \otimes \cdots \otimes I \otimes \ga \otimes I \otimes \cdots \otimes I$, with $\ga$ at the $i$-th of the $N$ places, and $A$ is an external electromagnetic potential.

The relativistic version of the wave function $\psi_t$ is $\psi_\Si:=\psi|_{\X_\Si}$, and the relativistic versions of the probability density $\rho_t$ and of the probability current $\j_{it}$ are:
\bg{eqnarray}
& & \rho_\Si(\si):=\overline \psi_\Si(\si) (\ga_1 \cdot n_\Si(\si_1)) \cdots (\ga_N \cdot n_\Si(\si_N)) \psi_\Si(\si);\\
& & j_{i \Si}(\si):=\overline \psi_\Si (\si) (\ga_1 \cdot n_\Si(\si_1)) \cdots \ga_i \cdots (\ga_N \cdot n_\Si(\si_N)) \psi_\Si(\si), \; \; i=1, \ldots, N,
\end{eqnarray}
where $\si = (\si_1, \ldots, \si_N) \in \X_\Si$ and $n_\Si(\si_i)$ is the future-oriented unit normal vector on $\Si$ at the point $\si_i \in \Si$. In \cite{hyper} it is proved that $\rho_\Si(\si) \geq 0$. Moreover, by applying the divergence theorem to the divergence-free tensor $\ops(x) \ga^{\mu_1}_1 \cdots \ga^{\mu_N}_N \psi(x)$ one can prove that the integral 
\bg{equation} \label{nr}
\int_{\X_\Si} \rho_\Si \, d^{\,3N}\si
\end{equation}
does not depend on $\Si$, and it will be assumed to be equal to 1. In regard to the probability current, one can prove that $j_{i \Si}(\si)$ is a future-oriented non space-like 4-vector \cite{hyper}.

An important fact of relativistic Bohmian mechanics is that, in general, no probability space $(\B, P)$ exists, where $\B \se \K$, such that 
\bg{equation}
P(B)= P_\Si(B_\Si) \tx{ for all } B \se \B \tx{ and all } \Si,
\end{equation}
where $P_\Si$ is the probability measure on $\X_\Si$ induced by $\rho_\Si$. This corresponds to the fact that ``Quantum equilibrium cannot hold in all Lorentz frames'' \cite{lorentz}. However, if a foliation is chosen, a relativistic version of the Bohmian space (\ref{stand}) can be obtained.

Let $\FF$ be a given foliation; the Bohmian trajectories relative to $\FF$ are the trajectories $k=(k_1, \ldots, k_N) \in \K$ satisfying 
\bg{equation} \label{guirel}
\dot k_i(\Si) \propto j_{i \Si}(k(\Si)) \tx{ for all } \Si \in \FF \tx{ and } i=1, \ldots, N,
\end{equation}
where $\dot k_i (\Si)$ is the future-oriented unit vector tangent to $k_i$ at the point $k_i \cap \Si$. Equation (\ref{guirel}) is the relativistic version of the guiding equation. Let $\B_\FF$ denote the set of the Bohmian trajectories defined in this way; one can prove that \cite{hyper}:
\bg{equation}
P_\Si(B_\Si)=P_{\Si'}(B_{\Si'}) \tx{ for all } B \se \B_\FF \tx{ and } \Si, \Si' \in \FF.
\end{equation}
This equation is the relativistic version of the equivariance equation (\ref{equi}). A probability measure $P_\FF$ on $\B_\FF$ is then naturally defined by
\bg{equation}
P_\FF(B):= P_\Si (B_\Si), \tx{ where } B \se \B_\FF \tx{ and } \Si \in \FF.
\end{equation}
The probability space $(\B_\FF, P_\FF)$ will be referred to as a \tit{relativistic Bohmian space}. 

Note that it may be that $\B_\FF \cap \B_{\FF'} \neq \es$ for $\FF \neq \FF'$. In fact, if $k$ is a Bohmian trajectory generated by a foliation $\FF$, and $\FF'$ is a foliation obtained by changing any leaf $\Si$ of $\FF$ somewhere or everywhere with the exception of a neighborhood of $k_1(\Si) \cup \ldots \cup k_N(\Si)$, then $\FF'$ also generates $k$. It will be assumed without proof that all the foliations generating $k$ can be constructed in this way. One easily see that this defines a \tit{synchronization} \cite{lorentz} for the $N$ world lines of $k$. In conclusion, there is not a one-to-one correspondence between trajectories and foliations, but any trajectory of $\cup_\FF \B_\FF$ is endowed with a synchronization.

In order to study the behavior of the space $(\B_\FF, P_\FF)$ under \Pe transformations, let us temporarily denote it by $(\B^\psi_\FF, P^\psi_\FF)$, where the dependence on the wave function is made explicit (the $\psi$-dependence will generally be omitted in the notation, and will be restored only when the transformation properties are considered). A \Pe transformation $g$ on $M$ acts naturally on any subset of $\K$ and on any foliation and acts on $\psi$ by means of a suitable representation $U_g$ of the \Pe group. One can prove that
\bg{equation} \label{cov}
g\B^\psi_\FF=\B^{U_g \psi}_{ g\FF} \tx{ and } P^\psi_\FF(B)=P^{U_g \psi}_{g\FF}(gB), \tx{ where } B \se \B^\psi_\FF.
\end{equation}
These equations show that the relativistic Bohmian spaces are covariant.

As pointed out in the introduction, any space $(\B_\FF, P_\FF)$ could be an empirically adequate model of the universe, in the sense that any foliation could be the preferred foliation (if a preferred foliation does exist). This implies in particular that the set of the physical trajectories of any space $(\B_\FF, P_\FF)$ is typical.

\section{The no-law model} \label{nl}

Let us assume the no-law hypothesis, i.e., there is no law for the foliations, neither dynamical nor statistical. The absence of a dynamical law implies that all the foliations are allowed\footnote{This point can be better understood by means of the following analogy: in non-relativistic Bohmian mechanics the set $\B$ defined in Section \ref{non} is the set of the trajectories which are dynamically allowed by the guiding equation. If the guiding equation is removed as a law, the set of the allowed trajectories becomes the entire kinematic space $\K$.}. As a consequence all the trajectories of the space
\bg{equation}
\B:=\cup_\FF \B_\FF
\en{equation}
are dynamically allowed, and therefore $\B$ becomes the sample space of the no-law model.

Let us define the typical subsets of $\B$ as follows: a set $B \se \B$ is typical if
\bg{equation} \label{ct}
P_\FF(B \cap \B_\FF) \app 1 \tx{ for all } \FF.
\en{equation}
This definition can be justified as follows. Let $\FFF$ denote the set of the foliations. According to the no-law hypothesis no probability measure is defined on $\FFF$. If instead a probability measure were defined, the typical sets of $\B$ could be easily defined. In fact, a probability measure $\mu$ on $\FFF$ induces the probability measure
\bg{equation}
P_\mu(B):=\int_\FFF P_\FF(B \cap \B_\FF) d\mu(\FF)
\en{equation}
on $\B$. It is easy to see that a set $B$ is typical according to $P_\mu$ for any $\mu$ if and only if the condition (\ref{ct}) is satisfied. The condition (\ref{ct}) therefore defines the typical sets independently from any probability measure on $\FFF$, and therefore it is a reasonable definition of typicality also in absence of any probability measure.

In order to express the condition (\ref{ct}) in a more synthetic way let us introduce the set function
\bg{equation} \label{sf}
P_*(B) :=\inf_\FF P_\FF(B \cap \B_\FF), \tx{ where } B \se \B.
\en{equation}
One can easily prove that 
\bg{equation}
P_*(\es)=0, \; P_*(\B)=1, \tx{ and } P_*(B) \leq P_*(B') \tx{ for } B \se B'.
\en{equation}
Moreover $P_*$ is superadditive, i.e., $P_*(B \cup B') \geq P_*(B) + P_*(B') \tx{ for } B \cap B'= \es$. This set function is very similar to the lower probability set function utilized in the theory of imprecise probability \cite{ip}. Since $P_*(B) \app 1 \Lra P_\FF(B \cap \B_\FF) \app 1 \; \forall \; \FF$, the typicality condition (\ref{ct}) can be expressed as
\bg{equation} \label{ct2}
P_*(B) \app 1.
\en{equation}

In conclusion, the no-law model is the space 
\bg{equation}
(\B,P_*).
\en{equation}
One can easily prove that the space $(\B^\psi,P_*^\psi)$ is covariant, that is:
\bg{equation} \label{li}
g\B^\psi=\B^{U_g \psi} \tx{ and } P_*^\psi(B)=P^{U_g \psi}_*(gB), \tx{ where } B \se \B^\psi.
\en{equation}

The actual trajectory of the universe is assumed to be a trajectory chosen at random from $\B$, where the random choice is governed by the set function $P_*$ according the typicality condition (\ref{ct2}). Since there is not a one-to-one correspondence between trajectories and foliations, the random choice of a trajectory from $\B$ does not imply that a foliation is also chosen, and therefore one cannot say that a particular foliation is realized (see also the remark below). Recall however that every trajectory of $\B$ is endowed with a synchronization.

The empirical adequacy of the space $(\B, P_*)$, i.e., the fact that the subset of the physical trajectories of this space is typical, descends trivially from the fact that the subset of the physical trajectories of any space $(\B_\FF, P_\FF)$ is typical.

\vspace{3mm}
In the derivation of the space $(\B,P_*)$ it has been implicitly assumed that the foliations are not part of the primitive ontology of the model \cite{common}. In other words, they are assumed to have the same ontological nature of the wave function or of the probability measure, and not that of the trajectories. This also makes acceptable the fact that no particular foliation is realized.

A different model is derived if instead the foliations are assumed to be part of the primitive ontology. In this case they have to be included in the sample space, which becomes:
\bg{equation}
\B':=\cup_\FF \{\FF\} \times \B_\FF.
\en{equation}
The typical subsets of $\B'$ are defined by the set function 
\bg{equation} 
P_*'(B):=\inf_\FF P_\FF(B_\FF),
\en{equation}
where $B \se \B'$ and $B_\FF:=\{k \in \K: (\FF, k) \in B\}$. Unlike the first model, in this model the random choice of an element from $\B'$ also includes the choice of a foliation, and therefore in this case a particular foliation is realized.

The two models are however very similar, both from the formal and from the conceptual point of view. For example, even if in the first model the random choice does not include a foliation, the fact that any trajectory of $\B$ is endowed with a synchronization makes the two models very similar in regard to their compatibility with relativity.

\section{Discussion} \label{disc}

Let us discuss first the following possible objection: also according to the no-law model the actual trajectory of the universe has a preferred synchronization, and this structure is conceptually very similar to a foliation. This essentially contradicts the title of the paper.

A possible reply is the following. Usually the elements composing a theory can be classified at two different levels, namely the law-level and the instance-level. For example, in a theory based on a differential equation, the differential equation is at the law-level, while the single solutions are at the instance-level; in a theory based on a probability space, the sample space and the probability measure are at the law-level, while the single outcomes are at the instance-level. Various properties and features of a theory can be classified analogously.

In the formulation of relativistic Bohmian mechanics proposed in \cite{bemade}, in which a single preferred foliation is derived from the universal wave function, the preferred foliation is certainly at the law-level. On the contrary, in the no-law model the preferred synchronization is certainly at the instance-level. I argue that a theory can be qualified as a theory with a preferred foliation/synchronization only if such a structure is defined at the law-level.

In order to support this conclusion consider the following example. Einstein's equations are at the law-level of general relativity, while single solutions of these equations are at the instance-level. It is well known that any solutions with an initial singularity (Big Bang) defines a preferred foliation, namely the foliation composed of the surfaces of constant time-like distance from the singularity. General relativity may therefore have preferred foliations at the instance-level, but certainly it cannot be qualified as a theory with a preferred foliation.

\vspace{3mm}
Let us now discuss the compatibility of the no-law model with relativity. The authors of \cite{bemade} distinguish between a Lorentz invariant theory and a fundamentally relativistic theory. Lorentz invariance is a well defined formal property which basically corresponds to equations (\ref{li}), and therefore the no-law model is Lorentz invariant. On the contrary, there is not an obvious criterion which states clearly if a theory is fundamentally relativistic or not, so that only intuitive considerations are possible here.

The fact that the actual trajectory of the universe is endowed with a preferred synchronization seems to be incompatible with fundamental relativity. On the other hand, the fact that the preferred synchronization is at the instance-level could reduce this incompatibility. The example of general relativity is also useful in this case: there are solutions of Einstein's equations which define a preferred foliation, but certainly general relativity is compatible with fundamental relativity.

\vspace{3mm}
Let us now address the main motivation for the no-law hypothesis, namely the empirical irrelevance of any law for the foliation. In fact, since the foliations are \tit{completely} unobservable, such a law cannot be even approximately empirically verified; moreover it can be ignored without any loss of predictive power, because any space $(\mathbb{B}_{\cal F}, P_{\cal F})$ could be an empirically adequate model of the universe.

Since empirical irrelevance is a criticism which is often (incorrectly) raised against the guiding equation of Bohmian mechanics, which is the dynamical law for the trajectories, it is useful to emphasize the difference between the two situations. First of all, unlike foliations, Bohmian trajectories are not completely unobservable. In fact only the microscopic details of a trajectory are unobservable, while its macroscopic structure determines the observable trajectories of the macroscopic bodies. As a consequence, even if we cannot be certain that the guiding equation is exact, we can at least verify it on the macroscopic level. Second, the role of the guiding equation is to define a set of trajectories, i.e., it participates in the definition of a model. On the contrary, the role of a law for the foliation is to select a model from an already existing class of empirically equivalent models, namely the class $\{(\B_\FF, P_\FF)\}_{\FF \in \FFF}$. For this reason, unlike the law for the foliation, the guiding equation cannot be ignored without a radical loss of predictive power. For example, if the guiding equation is ignored, the quasi-classical evolution of the macroscopic world can no longer be predicted.

\bg{thebibliography}{10}
\bibitem{common} Allori, V., Goldstein, S., Tumulka, R., Zangh\`i, N.: On the Common Structure of Bohmian Mechanics and the Ghirardi-Rimini-Weber Theory. Brit. J. Philos. Sci. 59, 353–389 (2008) arXiv:quant-ph/0603027

\bibitem{lorentz} Berndl, K., D\"urr, D., Goldstein, S., Zangh\`i, N.: Nonlocality, Lorentz Invariance, and Bohmian Quantum Theory. Phys. Rev. A 53, 2062-2073 (1996) arXiv:quant-ph/9510027

\bibitem{bohm} Bohm, D., Hiley, B.J.: The Undivided Universe. Routledge, New York (1993)

\bibitem{hyper} D\"urr, D., Goldstein, S., M\"unch-Berndl, K., Zangh\`i, N.: Hypersurface Bohm-Dirac models. Phys. Rev. A 60, 2729-2736 (1999) arXiv:quant-ph/9801070

\bibitem{bemade}  D\"urr, D., Goldstein, S., Norsen T., Struyve, W., Zangh\`i, N.: Can Bohmian mechanics be made relativistic? Proc. Roy. Soc. A 470, 20130699 (2013) arXiv:1307.1714

\bibitem{dbook2} D\"urr, D., Goldstein, S., Zangh\`i, N.: Quantum Physics without Quantum Philosophy. Springer-Verlag, Berlin (2013)

\bibitem{qe} D\"urr, D., Goldstein, S., Zangh\`i, N.: Quantum Equilibrium and the Origin of
Absolute Uncertainty. J. Statist. Phys. 67, 843–907 (1992) arXiv:quant-ph/0308039

\bibitem{dbook}  D\"urr, D., Teufel, S.: Bohmian Mechanics. Springer-Verlag, Berlin (2009)

\bibitem{gal} Galvan, B.: Generalization of the Born rule. Phys. Rev. A 78, 042113 (2008) arXiv:0806.4935

\bibitem{typ0} Goldstein, S.: Boltzmann's Approach to Statistical Mechanics. In: Bricmont, J., D\"urr, D., Galavotti, M. C., Ghirardi, G., Petruccione, F., Zangh\'i, N. (eds.): Chance in Physics: Foundations and Perspectives, pp. 39-54. Lecture notes in Physics 574. Springer-Verlag Berlin (2001) arXiv:cond-mat/0105242 

\bibitem{typ} Goldstein, S.: Typicality and Notions of Probability in Physics. In: Ben-Menahem, Y., Hemmo, M. (eds.): Probability in Physics, pp. 59-71. Springer Berlin (2012)

\bibitem{hol} Holland, P.R.: The quantum Theory of Motion. Cambridge University Press, Cambridge
(1993)

\bibitem{ip} Huber, P. J.: Robust Statistics. pp. 253-258. Wiley, New York (1981)

\bibitem{maud} Maudlin, T.: Space-time in the quantum world. In: Cushing, J.T., Fine, A., Goldstein, S. (eds.): Bohmian Mechanics and Quantum Theory: An Appraisal. Kluwer Academic, Dordrecht (1996)

\bibitem{cp} Shafer, G.: From Cournot's principle to market efficiency. In: Touffut, J. P. (ed.) Augustin Cournot: Modelling Economics. Edward Elgar (2007) http://www.glennshafer.com

\end{thebibliography}

\end{document}